# Electromagnetic fields in nonuniform disk-loaded waveguides

M.I. Ayzatsky[1], V.V.Mytrochenko
National Science Center Kharkov Institute of Physics and Technology (NSC KIPT),
610108, Kharkov, Ukraine

On the base of general approach we obtain some results that can be useful in the process of tuning of nonunifrom disc-loaded waveguides. Our consideration has shown that simple values that characterize the detuning of the cells can be introduced only for the disc-loaded waveguide with parameters that change very slow. In general case it is needed to conduct full numerical simulation of specific disc-loaded waveguide and obtain all necessary coupling coefficients. After that one can start the tuning process on the base of bead-pull field distribution measurements.

.

## 1 Introduction

Disc-loaded waveguides (DLW) have been heavily investigated both numerically and analytically over the past seven decades (see, for example, [1,2] and cited there literature). They have also been used, and continue to be used, in a variety of microwave devices such as linear accelerators [3,4], travelling-wave tube amplifiers, backward-wave oscillators [5], etc.

In physics of linear accelerators two important groups of DLWs are commonly distinguished: the constant gradient and the constant-impedance type. Characteristics of a homogenous structure with constant iris and cell-radii over the whole length (a constant-impedance structure) can be calculated with using Floquet theorem. There are several known reliable methods and computer codes for calculations of the infinitely periodic structures.

A structure with constant electric field on axis (a constant-gradient structure) can be designed by appropriate tapering iris- and cell-radius [6,7,8]. The design of such nonperiodic accelerator section is based on the assumption that the parameters of the section vary only slowly along the structure. In such an approach one essentially assumes the local field in the nonperiodic structure to be that of a corresponding periodic structure, with the local dimensions of the nonperiodic structure. However, there always arises the question of how good is this approximation for the nonperiodic wave structures.

Some nonuniform DLWs have greater tapering. They are used as injectors and sections for high current linacs [3,9,10]. There are also quasi constant gradient structures that have essentially nonuniform transition cells [11,12].

Usually the electromagnetic properties of accelerator components are calculated by computer codes that discretize Maxwell's equations. For long tapered disc-loaded waveguides, however, these methods would need the solution of extremely large algebraic equations. This is numerically difficult (or even impossible).

So, it is necessary to use non-grid-oriented methods to calculate the fields in the complete structure with realistic dimensions. Equivalent circuits are one possible technique. These are fast methods but the influence of the chosen model on the results is not negligible so that the results may be far away from the exact solution of Maxwell's equations. The mode matching technique is based on an exact formulation. In the chain matrix formulation this technique can be used for nonperiodic structures [13,14,15].

Usually, in the mode matching technique basic functions are chosen as the eigenwaves of circular waveguides [13,14,15]. Earlier we have developed approach that used the eigenmodes of circular cavities as the basic functions for calculation the properties of uniform DLW [16]. We have obtained exact infinite system of coupled equations which can be reduced by making some assumptions. Under such procedure we can receive more exact parameters of equivalent circuits by solving the appropriative algebraic systems. These parameters of equivalent circuits are functions both geometric sizes and frequency. Moreover, under such approach all used values have interpretation. We called this approach as coupled cavity model. Recently we extended that model on the case of nonuniform DLW. The detailed description of this nonuniform model will be given in another article.

[1] M.I. Aizatskyi, N.I.Aizatsky; aizatsky@kipt.kharkov.ua

Development of this model gives us possibility to look on the properties of the nonuniform DLW more deeply, especially on the methods that are used for tuning nonuniform DLWs [17,18,19,20,21,22,23,24]]. Some results of our investigation are presented in this article.

## 2 Electromagnetic fields in nonuniform disk-loaded waveguide

Let's consider a cylindrical nonuniform DLW (Fig.1). We will consider only axially symmetric fields with $E_z, E_r, H_\varphi$ components. Time dependence is $\exp(-i\,\omega t)$.

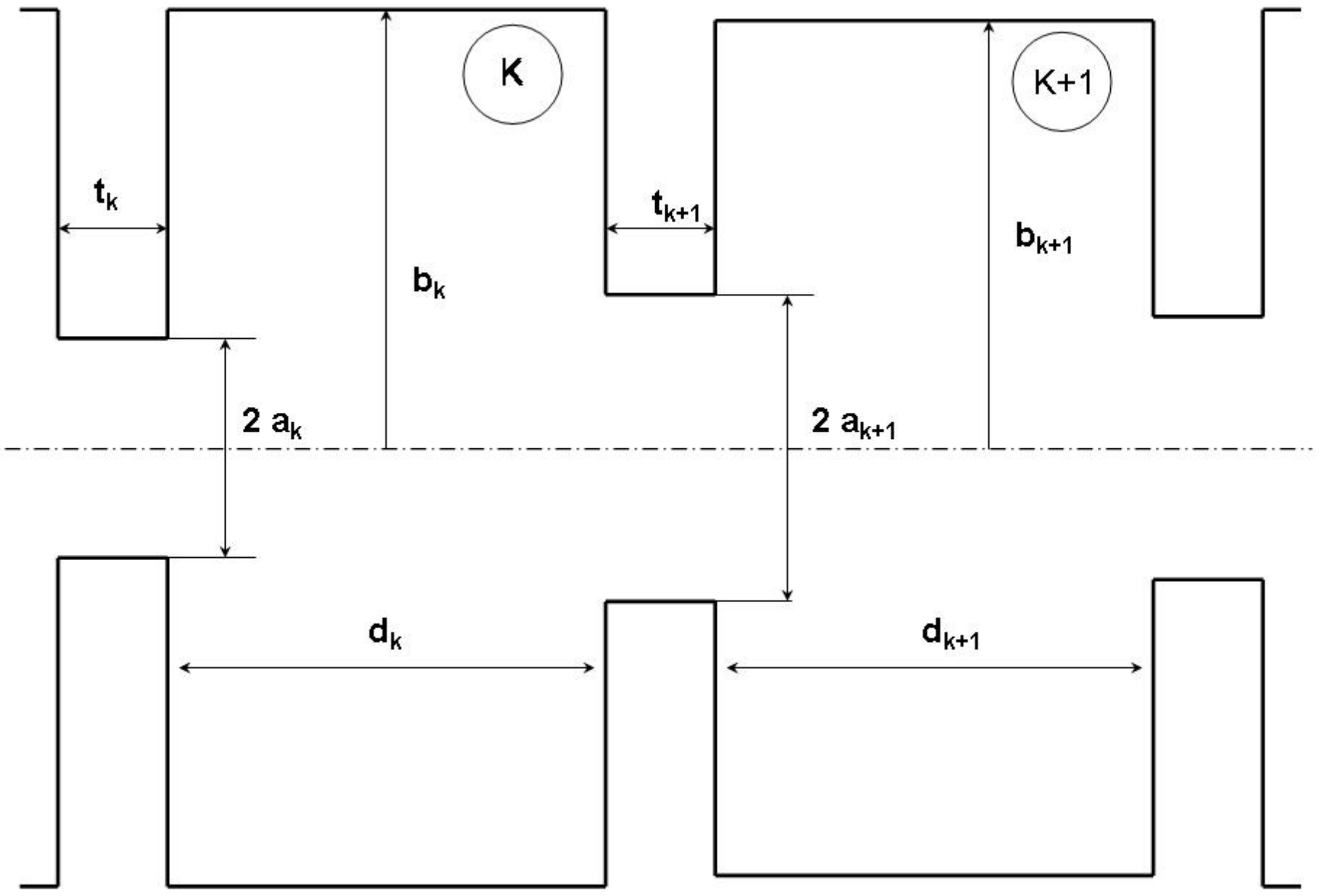

**Fig. 1**

We can divide the DLW volume into infinite number of different cylindrical volumes that are contiguous with each other over circle aria. In each large volume we expand the electromagnetic field with the short-circuit resonant cavity modes

$$\vec{E}^{(k)} = \sum_q e_{q,k}(t)\,\vec{E}_{q,k}(\vec{r}) \ , \tag{1.1}$$

$$\vec{H}^{(k)} = i\sum_q h_{q,k}(t)\,\vec{H}_{q,k}(\vec{r}) , \tag{1.2}$$

where $q = \{0,m,n\}$, and $\vec{E}_{q,k}$, $\vec{H}_{q,k}$ are the solution of homogenous Maxwell equations

$$\begin{aligned} rot\,\vec{E}_{q,k} &= i\,\omega_{q,k}\mu_0\vec{H}_{q,k} \ , \\ rot\,\vec{H}_{q,k} &= -i\,\omega_{q,k}\varepsilon_0\vec{E}_{q,k} \end{aligned} \tag{1.3}$$

with boundary condition $\vec{E}_\tau = 0$ on the metal surface. In each small volume we expand the electromagnetic field with the waveguide modes. Using the relevant boundary conditions after some manipulations we can obtain such infinite set of coupled relations

$$\left(\omega_{q,k}^2-\omega^2\right)e_{q,k}=\omega_{q,k}^2\sum_{j=-\infty}^{\infty}e_{010,j}\alpha_{q,k,j} \quad , \tag{1.4}$$

where $\alpha_{q,k,j}$ are some coefficients that depend on both the frequency $\omega$ and geometrical sizes of all volumes. There are an infinite set of infinite linear equations which solutions define these $\alpha_{q,k,j}$ coefficients. Let's note that relations (1.4) are the exact ones.

It follows from (1.4) that for finding the amplitudes of the main (010) mode we have to solve a system of coupled equations

$$\left\{\omega_{010,k}^2\left(1-\alpha_{010,k,k}\right)-\omega^2\right\}e_{010,k}=\omega_{010,k}^2\sum_{j=-\infty,\,j\neq k}^{\infty}e_{010,j}\alpha_{010,k,j}\,. \tag{1.5}$$

Amplitudes of other modes ($q\neq(010)$) can be find by summing the relevant series

$$e_{q,k}=\frac{\omega_{q,k}^2}{\left\{\omega_{q,k}^2\left(1-\alpha_{q,k,k}\right)-\omega^2\right\}}\sum_{j=-\infty,\,j\neq k}^{\infty}e_{010,j}\alpha_{q,k,j} \tag{1.6}$$

It can be shown that for large interval of geometric sizes in the right hand side of (1.5) we can neglect all terms except three ones. Denote $\bar{A}_k=e_{010,k}$ (complex amplitudes), $\alpha_{k,j}=\alpha_{010,k,j}$ and introduce loss, then the equations (1.5) can be rewritten as

$$\left[\omega_k^2(1-\alpha_{k,k})-i\frac{\omega_k\omega}{Q_k}-\omega^2\right]\bar{A}_k=\omega_k^2\alpha_{k,k+1}\bar{A}_{k+1}+\omega_k^2\alpha_{k,k-1}\bar{A}_{k-1}\,, \tag{1.7}$$

or

$$Z_k\bar{A}_k=\alpha_{k,k-1}\bar{A}_{k-1}+\alpha_{k,k+1}\bar{A}_{k+1}\,, \tag{1.8}$$

where

$$Z_k=\left[1-\alpha_{k,k}-\frac{\omega^2}{\omega_k^2}-i\frac{\omega}{\omega_kQ_k}\right]=\operatorname{Re}Z_k-i\varepsilon_k\,. \tag{1.9}$$

Suppose that for $\omega=\omega_0$ and some geometry ($Z_k=Z_k^{(0)}$) there is a certain (forward) amplitude distribution

$$\bar{A}_k^{(0)}=\bar{A}_{k-1}^{(0)}R_{k,k-1}^{(0)}\exp(i\varphi_k^{(0)})\,, \tag{1.10}$$

where $R_{k,k-1}^{(0)}$ are the real values.

Note that in general case electromagnetic field with amplitude distribution (1.10) can not be considered as a wave (except the uniform case). Indeed, such distribution can be true only for one frequency and in some cases it can be impossible to define the group velocity[2].

From (1.8) and (1.10) it follows

$$\bar{A}_{k+1}^{(0)}=\frac{1}{\alpha_{k,k+1}^{(0)}}\left(Z_k^{(0)}-\frac{\alpha_{k,k-1}^{(0)}}{R_{k,k-1}^{(0)}}\exp(-i\varphi_k^{(0)})\right)\bar{A}_k^{(0)}=\bar{A}_k^{(0)}\exp(i\varphi_{k+1}^{(0)})R_{k+1,k}^{(0)}\,. \tag{1.11}$$

Therefore, under $\omega=\omega_0$ and some geometry such relations is true

$$Z_k^{(0)}=\frac{\alpha_{k,k-1}^{(0)}}{R_{k,k-1}^{(0)}}\exp(-i\varphi_k^{(0)})+\exp(i\varphi_{k+1}^{(0)})R_{k+1,k}^{(0)}\alpha_{k,k+1}^{(0)}\,. \tag{1.12}$$

For given $R_{k,k-1}^{(0)}$, $\varphi_k^{(0)}$ these equations determine the geometry of DLW.

[2] In a periodic structure, at a given frequency and single mode operation, the electromagnetic wave is characterized by a single wavenumber *k* and quantities like phase velocity, group velocity. In principle, if the structure is no longer periodic the field can not be represented by a single wavenumber and group velocity. The simplest example is the matched (on one frequency) connection of two different DLWs. It is impossible to define the group velocity for this case. We will use term electromagnetic field (forward and backward) instead of electromagnetic wave.

The equations (1.8) are the second order difference equation. So, there must two solutions.

When $Q_k \to \infty$ the second solution is a complex conjugate one to (1.10), so, it is conveniently to seek the second solution in the form

$$\tilde{\tilde{A}}_k^{(0)} = \tilde{\tilde{A}}_{k-1}^{(0)} \tilde{R}_{k,k-1}^{(0)} \exp(-i\phi_k^{(0)}) . \quad (1.13)$$

When $Q_k \to \infty$ $\tilde{R}_{k,k-1}^{(0)} \to R_{k,k-1}^{(0)}$; $\phi_k^{(0)} \to \varphi_k^{(0)}$.

$\tilde{R}_{k,k-1}^{(0)}$, $\phi_k^{(0)}$ satisfy the equations

$$Z_k^{(0)} = \frac{\alpha_{k,k-1}^{(0)}}{\tilde{R}_{k,k-1}^{(0)}} \exp(i\phi_k^{(0)}) + \exp(-i\phi_{k+1}^{(0)}) \tilde{R}_{k+1,k}^{(0)} \alpha_{k,k+1}^{(0)} . \quad (1.14)$$

Consider a case of the uniform DLW (a constant-impedance structure) ($\phi_k^{(0)} = \varphi$, $R_{k,k-1}^{(0)} = const$, $\tilde{R}_{k,k-1}^{(0)} = const$, $\alpha_{k,k-1}^{(0)} = const$). From (1.12) it follows

$$R^{(0)} = -\frac{\varepsilon^{(0)}}{2\alpha^{(0)}\sin(\varphi)} \pm \sqrt{\left(\frac{\varepsilon^{(0)}}{2\alpha^{(0)}\sin(\varphi)}\right)^2 + 1} \approx 1 - \frac{\varepsilon^{(0)}}{2\alpha^{(0)}\sin(\varphi)} + \frac{1}{2}\left(\frac{\varepsilon^{(0)}}{2\alpha^{(0)}\sin(\varphi)}\right)^2 \quad (1.15)$$

$$\tilde{R}^{(0)} = \frac{\varepsilon^{(0)}}{2\alpha^{(0)}\sin(\varphi)} \pm \sqrt{\left(\frac{\varepsilon^{(0)}}{2\alpha^{(0)}\sin(\varphi)}\right)^2 + 1} \approx 1 + \frac{\varepsilon^{(0)}}{2\alpha^{(0)}\sin(\varphi)} + \frac{1}{2}\left(\frac{\varepsilon^{(0)}}{2\alpha^{(0)}\sin(\varphi)}\right)^2 \quad (1.16)$$

For the case of the constant-gradient structure ($R_{k,k-1}^{(0)} = 1$ и $\phi_k^{(0)} = \varphi$) we have

$$Z_k^{(0)} = \alpha_{k,k-1}^{(0)} \exp(-i\varphi) + \alpha_{k,k+1}^{(0)} \exp(i\varphi) \quad (1.17)$$

$$-\varepsilon_k^{(0)} = \left(\alpha_{k,k+1}^{(0)} - \alpha_{k,k-1}^{(0)}\right)\sin\varphi \quad (1.18)$$

$$\alpha_{k,k+1}^{(0)} = \alpha_{k,k-1}^{(0)} - \frac{\varepsilon_k^{(0)}}{\sin\varphi} \quad (1.19)$$

$$\operatorname{Re} Z_k^{(0)} = \left(-\frac{\varepsilon_k^{(0)}}{\sin\varphi} + 2\alpha_{k,k-1}^{(0)}\right)\cos\varphi = \left(\frac{\varepsilon_k^{(0)}}{\sin\varphi} + 2\alpha_{k,k+1}^{(0)}\right)\cos\varphi \quad (1.20)$$

For backward distribution (1.13) from (1.14) we can obtain

$$\tilde{R}_{k,k-1}^{(0)} = \frac{\sin\varphi \sin(\phi_k^{(0)} + \phi_{k+1}^{(0)})\alpha_{k,k-1}^{(0)}}{\alpha_{k,k+1}^{(0)} \sin\phi_{k+1}^{(0)} \sin 2\varphi - \sin\left(\varphi - \phi_{k+1}^{(0)}\right)\varepsilon_k^{(0)}} . \quad (1.21)$$

$$\tilde{R}_{k+1,k}^{(0)} = \frac{\alpha_{k,k-1}^{(0)} \sin 2\varphi \sin\phi_k^{(0)} + \sin\left(\varphi - \phi_k^{(0)}\right)\varepsilon_k^{(0)}}{\sin\varphi \sin(\phi_k^{(0)} + \phi_{k+1}^{(0)})\alpha_{k,k+1}^{(0)}} . \quad (1.22)$$

We will seek the solution in the form

$$\phi_k^{(0)} = \varphi + \delta_k . \quad (1.23)$$

We will suppose bellow that $\varepsilon_k^{(0)} << 1$ and $\delta_k << \varphi$. Then

$$\tilde{R}_{k,k-1}^{(0)} \approx \frac{\alpha_{k,k-1}^{(0)}}{\alpha_{k,k+1}^{(0)}} \left\{1 + ctg 2\varphi(\delta_k + \delta_{k+1}) - ctg\varphi\delta_{k+1}\right\} . \quad (1.24)$$

$$\tilde{R}_{k+1,k}^{(0)} \approx \frac{\alpha_{k,k-1}^{(0)}}{\alpha_{k,k+1}^{(0)}} \left\{1 - ctg 2\varphi(\delta_k + \delta_{k+1}) + ctg\varphi\delta_k\right\} . \quad (1.25)$$

The right hand part of the equality (1.24) under change $k$ to $k+1$ must coincide with the right hand part of the equality (1.25)

$$\frac{\alpha_{k,k-1}^{(0)}}{\alpha_{k,k+1}^{(0)}} \left\{1 - ctg 2\varphi(\delta_k + \delta_{k+1}) + ctg\varphi\delta_k\right\} = \frac{\alpha_{k+1,k}^{(0)}}{\alpha_{k+1,k+2}^{(0)}} \left\{1 + ctg 2\varphi(\delta_{k+1} + \delta_{k+2}) - ctg\varphi\delta_{k+2}\right\} . \quad (1.26)$$

From this expression we can obtain such difference equation for $\delta_k$

$$\delta_{k+2}-\delta_{k+1}2\cos 2\varphi+\delta_k=2\cos\varphi\left(\frac{\varepsilon_{k+1}^{(0)}}{\alpha_{k+1,k}^{(0)}}-\frac{\varepsilon_k^{(0)}}{\alpha_{k,k-1}^{(0)}}\right). \quad (1.27)$$

The right hand part of this equation is proportional to $\varepsilon_k^{(0)2}$, therefore the solution will be big-oh of $\varepsilon_k^{(0)2}$, too. Therefore, we can neglect $\delta_k,\delta_{k+1}$ in (1.24), (1.25) and write

$$\phi_k^{(0)}\approx\varphi. \quad (1.28)$$

$$\tilde{R}_{k,k-1}^{(0)}\approx\frac{\alpha_{k,k-1}^{(0)}}{\alpha_{k,k+1}^{(0)}}\approx 1+\frac{\varepsilon_k^{(0)}}{\alpha_{k,k-1}^{(0)}\sin\varphi}. \quad (1.29)$$

## 3 Characterization of detuned cells

Let's consider the DLW with geometry close to some DLW that support $\overline{A}_k^{(0)}=\overline{A}_{k-1}^{(0)}R_{k,k-1}^{(0)}\exp(i\varphi_k^{(0)})$ distribution. Suppose that we can measure the complex amplitude $\overline{E}_k^{(m)}$ of the electromagnetic field in the center of each cavity and find the complex amplitude of $E_{010}$ mode $\overline{A}_k^{(m)}$:

$$\overline{A}_k^{(m)}=\chi_k\overline{E}_k^{(m)}. \quad (1.30)$$

Then under the assumption of only adjacent interaction these amplitudes have to satisfy the equation

$$Z_k\overline{A}_k^{(m)}=\alpha_{k,k-1}\overline{A}_{k-1}^{(m)}+\alpha_{k,k+1}\overline{A}_{k+1}^{(m)} \quad (1.31)$$

Rewrite it as

$$Z_k=\frac{\alpha_{k,k-1}^{(0)}\overline{A}_{k-1}^{(m)}+\alpha_{k,k+1}^{(0)}\overline{A}_{k+1}^{(m)}}{\overline{A}_k^{(m)}}+\left(\alpha_{k,k-1}-\alpha_{k,k-1}^{(0)}\right)\frac{\overline{A}_{k-1}^{(m)}}{\overline{A}_k^{(m)}}+\left(\alpha_{k,k+1}-\alpha_{k,k+1}^{(0)}\right)\frac{\overline{A}_{k+1}^{(m)}}{\overline{A}_k^{(m)}} \quad (1.32)$$

Let's denote

$$Z_k^{(3.m)}=\frac{\alpha_{k,k-1}^{(0)}\overline{A}_{k-1}^{(m)}+\alpha_{k,k+1}^{(0)}\overline{A}_{k+1}^{(m)}}{\overline{A}_k^{(m)}} \quad (1.33)$$

then

$$Z_k=Z_k^{(3,m)}+\left(\alpha_{k,k-1}-\alpha_{k,k-1}^{(0)}\right)\frac{\overline{A}_{k-1}^{(m)}}{\overline{A}_k^{(m)}}+\left(\alpha_{k,k+1}-\alpha_{k,k+1}^{(0)}\right)\frac{\overline{A}_{k+1}^{(m)}}{\overline{A}_k^{(m)}} \quad (1.34)$$

Since for $Z_k\to Z_k^{(0)}$ differences $\left(\alpha_{k,k-1}-\alpha_{k,k-1}^{(0)}\right)\to 0$, $\left(\alpha_{k,k+1}-\alpha_{k,k+1}^{(0)}\right)\to 0$, then we can introduce such parameters that characterize the detuning of the cell

$$F_k^{(3,m)}=\frac{Z_k^{(3,m)}-Z_k^{(0)}}{Z_k^{(0)}} \quad (1.35)$$

If we calculate $Z_k^{(0)}$, $\alpha_{k,k-1}^{(0)},\alpha_{k,k+1}^{(0)}$, $\chi_k$ and measure $\overline{A}_k^{(m)},\overline{A}_{k-1}^{(m)}$, $\overline{A}_{k+1}^{(m)}$, we'll be able to find $F_k^{(3,m)}$.

Changing the geometry of $k$ cell in such a way that $\left|F_k^{(3,m)}\right|\to 0$, we obtain $Z_k=Z_k^{(0)}$, i.e. we obtain tuned cell under arbitrary parameters of other cells and reflections from couplers.

Consider the constant-gradient structure ($R_{i+1,i}=1$) with slow tapering ($\varphi\neq\pi/2$, $\alpha_{i,i-1}\approx\alpha_{i,i+1}$, $\chi_k\approx const$)

$$Z_k^{(0)}=\frac{\alpha_{k,k-1}^{(0)}}{R_{k,k-1}^{(0)}}\exp(-i\varphi_k^{(0)})+\exp(i\varphi_{k+1}^{(0)})R_{k+1,k}^{(0)}\alpha_{k,k+1}^{(0)}\approx 2\alpha_{k,k-1}^{(0)}\cos\varphi. \quad (1.36)$$

$$Z_k^{(3,m)} \approx \alpha_{k,k-1}^{(0)} \frac{\overline{A}_{k-1}^{(m)} + \overline{A}_{k+1}^{(m)}}{\overline{A}_k^{(m)}} \tag{1.37}$$

Then

$$F_k^{(3,m)} = \frac{Z_k^{(3,m)}}{Z_k^{(0)}} - 1 \approx \frac{\overline{A}_{k-1}^{(m)} + \overline{A}_{k+1}^{(m)} - 2\cos\varphi \overline{A}_k^{(m)}}{\overline{A}_k^{(m)} 2\cos\varphi}. \tag{1.38}$$

For $\varphi = 2\pi/3$

$$F_i^{(3,m)} = -\frac{\overline{A}_{i-1}^{(m)} + \overline{A}_{i+1}^{(m)} + \overline{A}_i^{(m)}}{\overline{A}_i^{(m)}} = \frac{i\sqrt{3}S_i \exp(i\theta_i)}{\overline{A}_i^{(m)}}, \tag{1.39}$$

where $S_i \exp(i\theta_i)$ are coefficients that were introduced in the article [17] and which should to be decreased in the process of tuning.

$$iS_i \exp(i\theta_i) = F_i^{(3,m)} \overline{A}_i^{(m)} / \sqrt{3}, \tag{1.40}$$

Imaginary part of $Z_k$ is small ($\sim 1/Q_k$, see (1.9)), therefore the imaginary part of $F_i^{(3,m)}$ is small, too, and the authors of the article [17] dealt only with $\mathrm{Re}\{iS_i \exp(i\theta_i)\}$.

In the case of the uniform DLW (a constant-impedance structure) ($\phi_i = \varphi \neq \pi/2$, $R_{i+1,i} = R$, $\alpha_{i,i+1} = \alpha$)

$$F_i^{(3,m)} = \frac{\overline{A}_{i-1}^{(m)} + \overline{A}_{i+1}^{(m)}}{\overline{A}_i^{(m)}} \frac{R}{\exp(i\varphi)R^2 + \exp(-i\varphi)} - 1. \tag{1.41}$$

As $R \approx 1 - \dfrac{\mathrm{Im}\,Z}{2\alpha\sin(\varphi)}$, then if $\left|\dfrac{\mathrm{Im}\,Z}{2\alpha\sin(\varphi)}\right| << 1$

$$F_k^{(3,m)} \approx \frac{\overline{A}_{k-1}^{(m)} + \overline{A}_{k+1}^{(m)} - 2\cos\varphi \overline{A}_k^{(m)}}{\overline{A}_k^{(m)} 2\cos\varphi}. \tag{1.42}$$

It coincides with expression (1.38).

For large apertures and small phase velocities for correct quantitative description we have to use five summands in the right hand side of equation (1.5)

$$Z_k \overline{A}_k = \alpha_{k,k+1}\overline{A}_{k+1} + \alpha_{k,k+2}\overline{A}_{k+2} + \alpha_{k,k-1}\overline{A}_{k-1} + \alpha_{k,k-2}\overline{A}_{k-2}, \tag{1.43}$$

Then we can introduce more complicated parameters that characterize the detuning of the cell

$$F_k^{(5,m)} = \frac{Z_k^{(5,m)}}{Z_k^{(0)}} - 1. \tag{1.44}$$

where

$$Z_k^{(5,m)} = \frac{\alpha_{k,k-2}^{(0)}\overline{A}_{k-2}^{(m)} + \alpha_{k,k-1}^{(0)}\overline{A}_{k-1}^{(m)} + \alpha_{k,k+1}^{(0)}\overline{A}_{k+1}^{(m)} + \alpha_{k,k+2}^{(0)}\overline{A}_{k+2}^{(m)}}{\overline{A}_k^{(m)}}. \tag{1.45}$$

$$\begin{aligned} Z_k^{(0)} &= \alpha_{k,k+1}^{(0)} \exp(i\varphi_{k+1}) R_{k+1,k} + \alpha_{k,k+2}^{(0)} R_{k+2,k+1} R_{k+1,k} \exp(i\varphi_{k+2} + i\varphi_{k+1}) + \\ &+ \frac{\alpha_{k,k-1}^{(0)}}{R_{k,k-1}} \exp(-i\varphi_k) + \frac{\alpha_{k,k-2}^{(0)}}{R_{k,k-1}R_{k-1,k-2}} \exp(-i\varphi_k - i\varphi_{k-1}) \end{aligned} \tag{1.46}$$

We have already noted that introduced above parameters that characterize the detuning of cells are correct in the case when parameters of other cell do not effect on their magnitude.

To study quality of introduced parameters from this point of view we calculated these parameters for different DLWs. We used the developed coupled cavity model for calculation the necessary coefficients. Consideration was carried out for frequency $f$ = 2856 MHz.

The first DLW represents the matched (on one frequency $f$ = 2856 MHz) connection of two different uniform DLWs with using one transition cell. Its parameters are given in Table 1 (disc thickness $t = 0.4$ cm). Transition cell sizes (a(10) and b(10)) were chosen by making the reflection coefficient small enough ($\left|\Gamma\right| < 10^{-3}$). Results of calculations that presented in Fig. 2-Fig. 6 show that simplified description ((1.39)) can be used in tuning procedure.

The second DLW (Table 2) represents the matched (on one frequency $f$ = 2856 MHz) connection of two different uniform DLWs with using one transition cell. Sizes of the second DLW change more steeply. Transition cell sizes (a(10) and b(10)) were chosen by the same method ($\left|\Gamma\right| < 3\times10^{-3}$). Results of calculations that presented in Fig. 7-Fig. 11show that for relatively fast cell size changes three point approach gives correct results only in the case of using full description (see (1.35)). Simplified description ((1.39)) gives some mistakes in considered case. Five point approach gives the same results as three point one.

But for the case of large apertures and small cell lengths (the third DLW, Table 3[3], small phase velocities) only five point approach gives correct results (see Fig. 12- Fig. 16).

**Table 1**

| N cell | a(cm) | b(cm) | d(cm) |
|---|---|---|---|
| 1 | 1.4 | 4.1686 | 3.0989 |
| 2 | 1.4 | 4.1686 | 3.0989 |
| 3 | 1.4 | 4.1686 | 3.0989 |
| 4 | 1.4 | 4.1686 | 3.0989 |
| 5 | 1.4 | 4.1686 | 3.0989 |
| 6 | 1.4 | 4.1686 | 3.0989 |
| 7 | 1.4 | 4.1686 | 3.0989 |
| 8 | 1.4 | 4.1686 | 3.0989 |
| 9 | 1.4 | 4.1686 | 3.0989 |
| 10 | 1.3964 | 4.1667 | 3.0989 |
| 11 | 1.39 | 4.1658 | 3.0989 |
| 12 | 1.39 | 4.1658 | 3.0989 |
| 13 | 1.39 | 4.1658 | 3.0989 |
| 14 | 1.39 | 4.1658 | 3.0989 |
| 15 | 1.39 | 4.1658 | 3.0989 |
| 16 | 1.39 | 4.1658 | 3.0989 |
| 17 | 1.39 | 4.1658 | 3.0989 |
| 18 | 1.39 | 4.1658 | 3.0989 |
| 19 | 1.39 | 4.1658 | 3.0989 |
| 20 | 1.39 | 4.1658 | 3.0989 |
| 21 | 1.39 | 4.1658 | 3.0989 |
| 22 | 1.39 | 4.1658 | 3.0989 |
| 23 | 1.39 | 4.1658 | 3.0989 |
| 24 | 1.39 | 4.1658 | 3.0989 |
| 25 | 1.39 | 4.1658 | 3.0989 |

[3] Cell radii were chosen with using special 6 resonator stack [10,11]. Tuning accuracy was ~50 kHz.

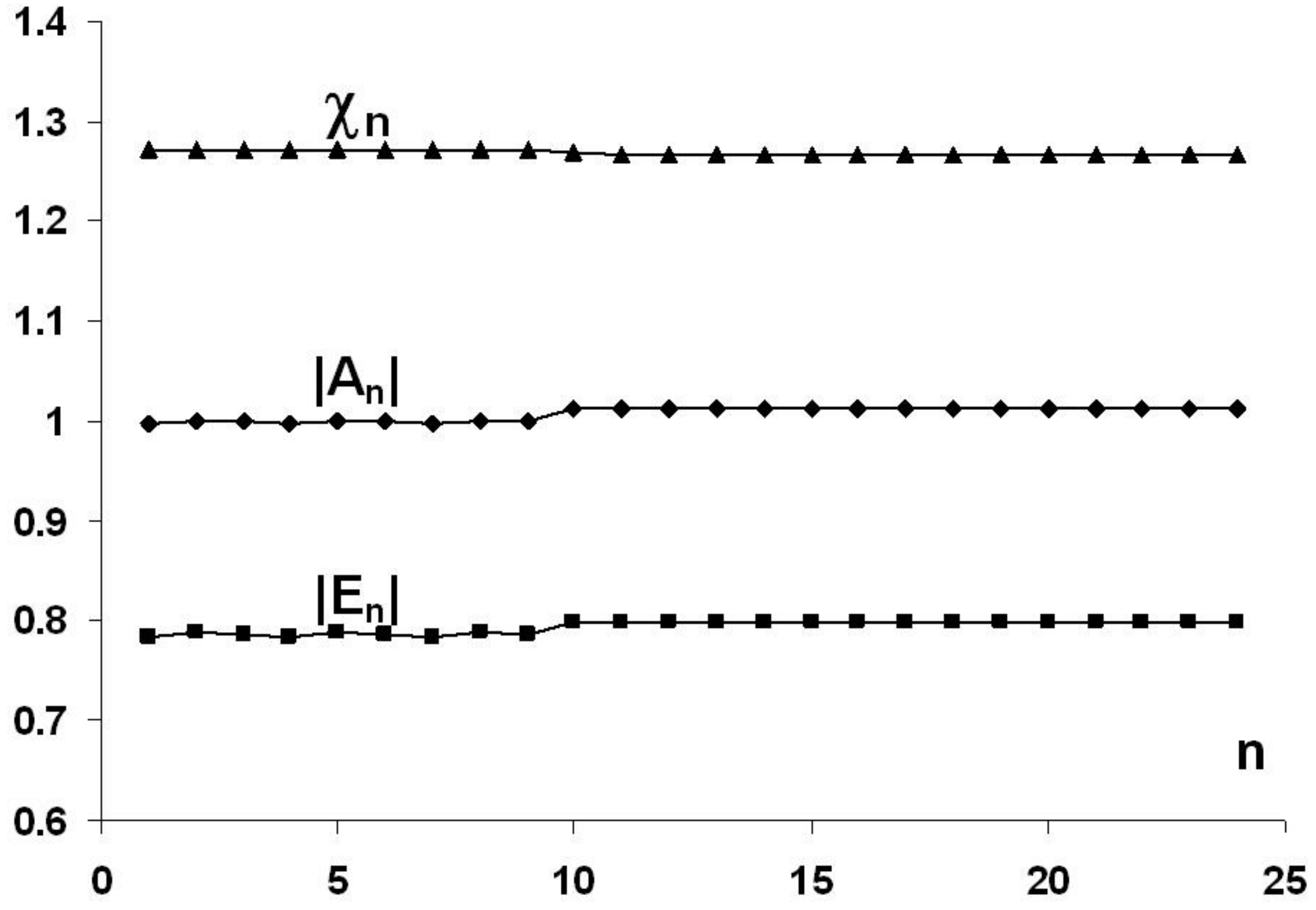

**Fig. 2 Distributions of longitudinal electric field in the cell centers ($|E_n|$), amplitudes of $E_{010}$ mode ($|A_n|$) and $\chi_n = |A_n / E_n|$**

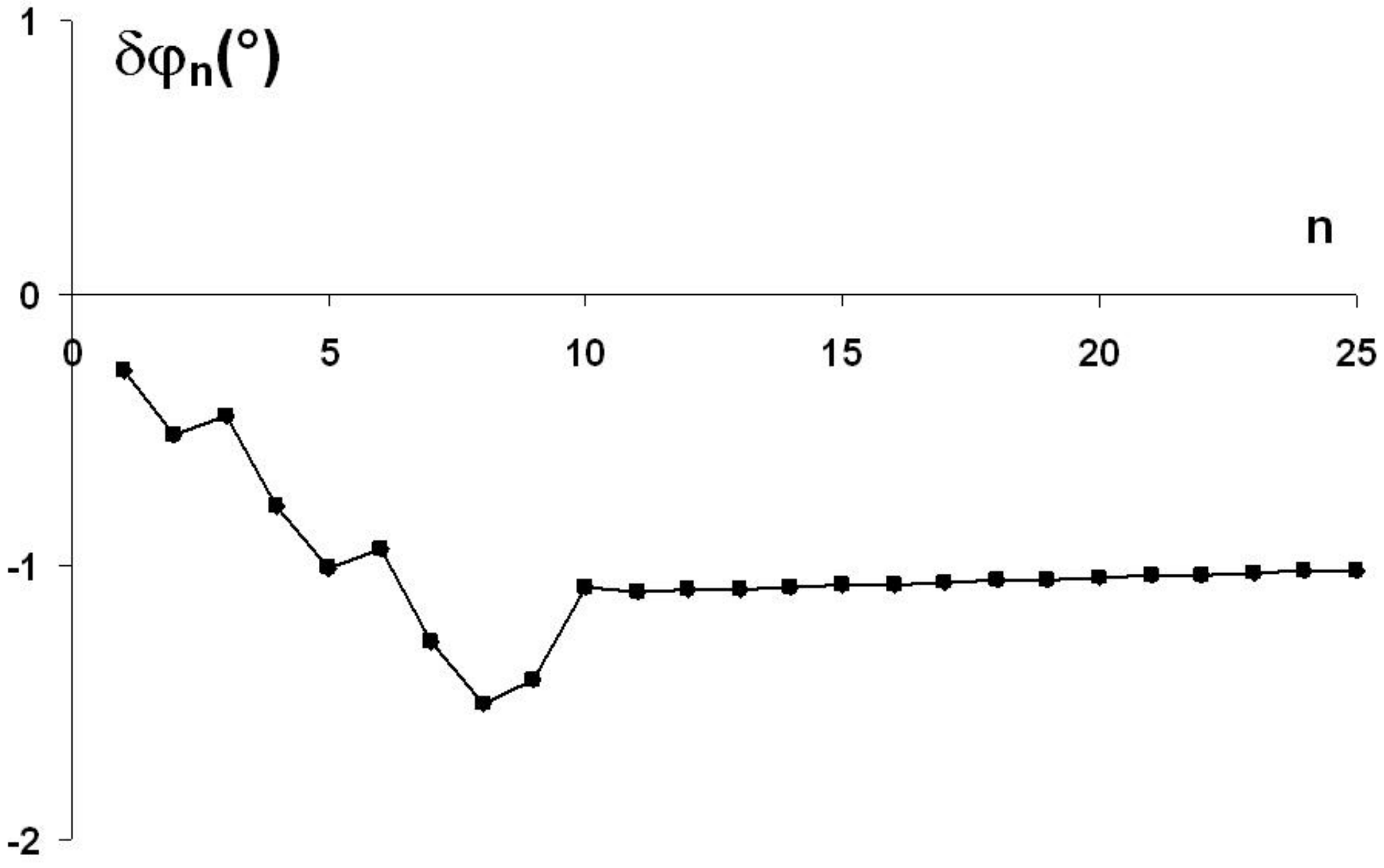

**Fig. 3 Distribution of phase deviations ($\varphi_0 = 2\pi/3$) of the longitudinal electric field in the cell centers and phase deviations of amplitude of $E_{010}$ mode**

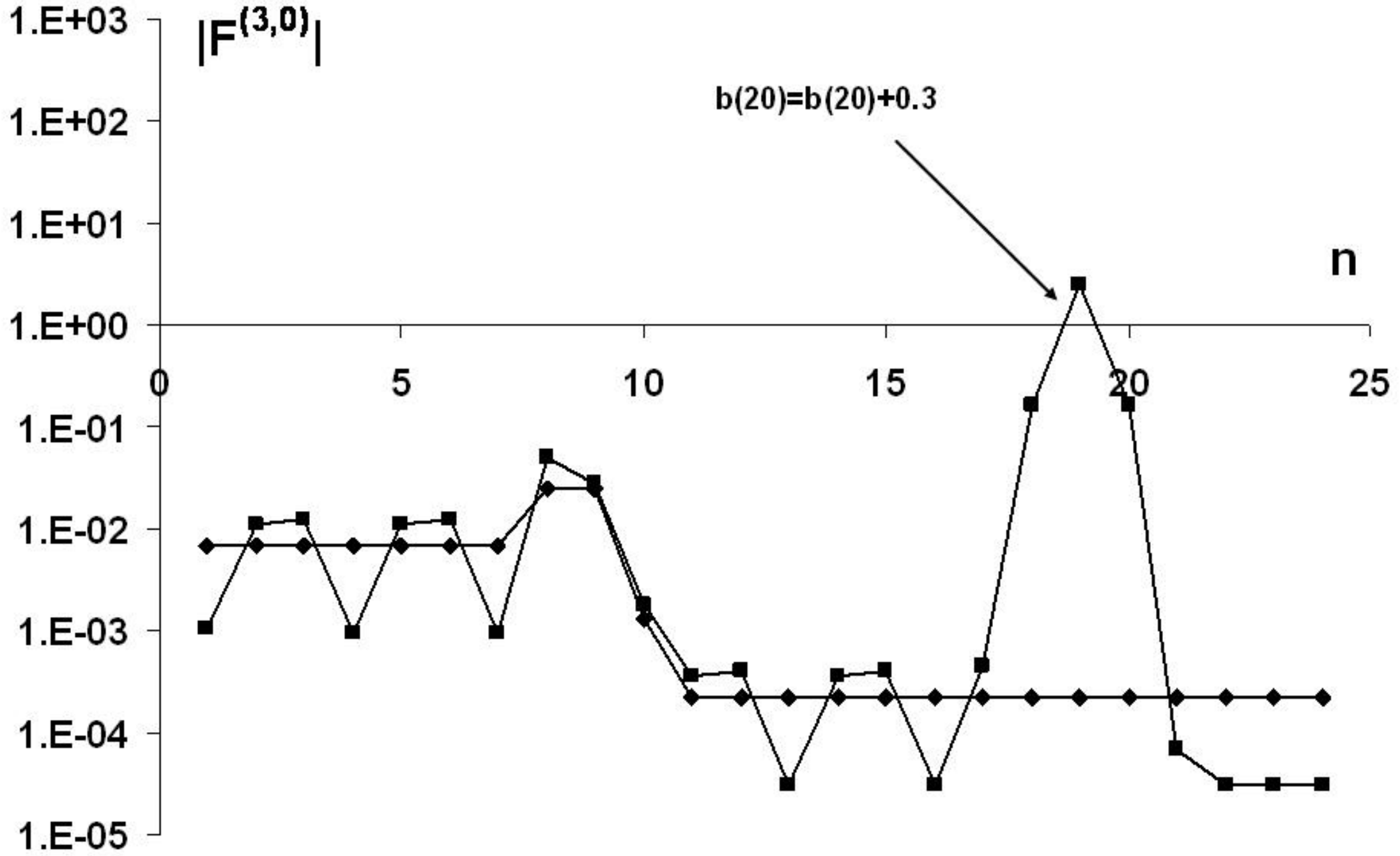

**Fig. 4 Detuning parameters that were calculated with using expression (1.39)**

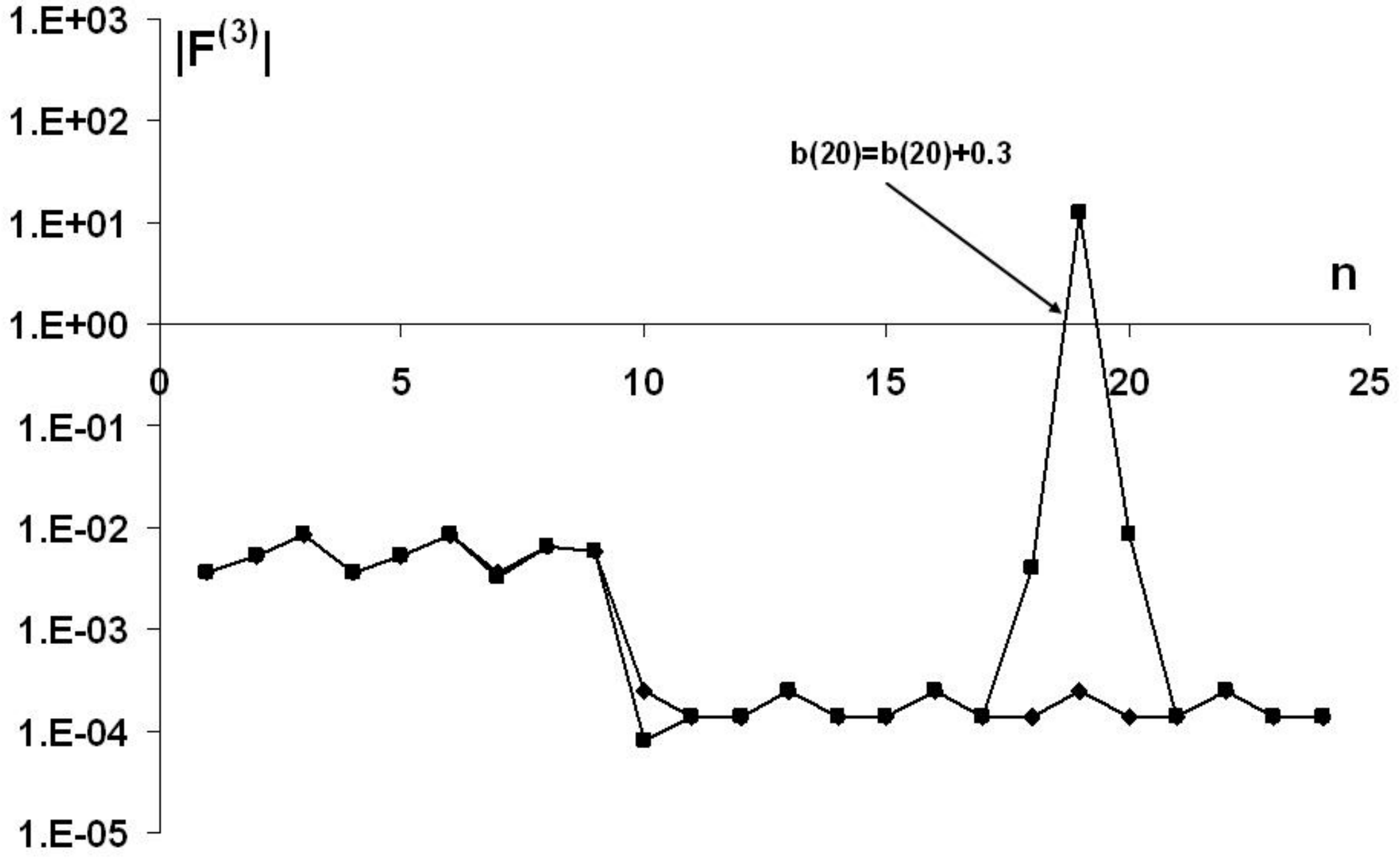

**Fig. 5 Detuning parameters that were calculated with using expression (1.35)**

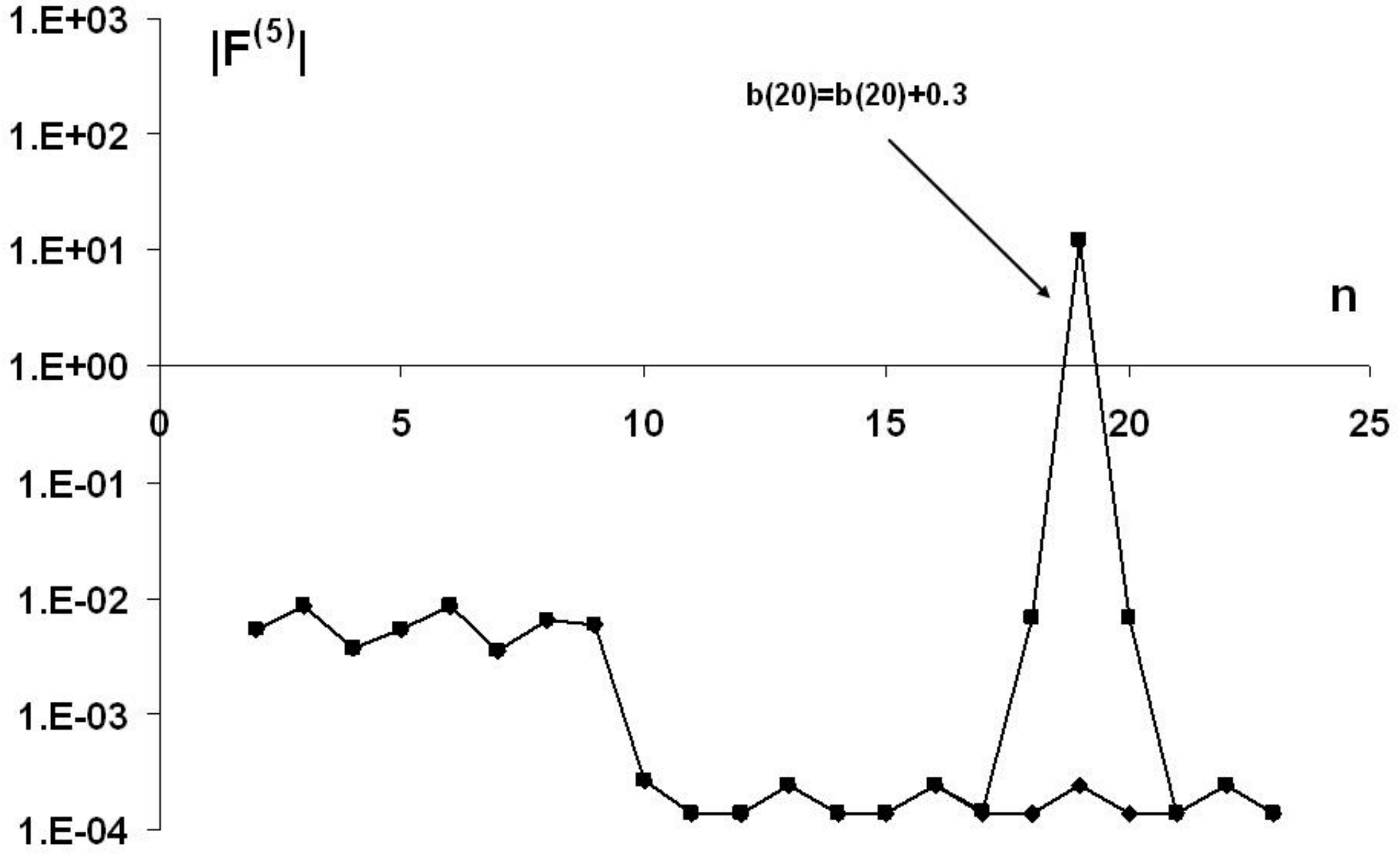

**Fig. 6 Detuning parameters that were calculated with using expression (1.44)**

**Table 2**

| N cell | a(cm) | b(cm) | d(cm) |
|---|---|---|---|
| 1 | 1.4 | 4.1686 | 3.0989 |
| 2 | 1.4 | 4.1686 | 3.0989 |
| 3 | 1.4 | 4.1686 | 3.0989 |
| 4 | 1.4 | 4.1686 | 3.0989 |
| 5 | 1.4 | 4.1686 | 3.0989 |
| 6 | 1.4 | 4.1686 | 3.0989 |
| 7 | 1.4 | 4.1686 | 3.0989 |
| 8 | 1.4 | 4.1686 | 3.0989 |
| 9 | 1.4 | 4.1686 | 3.0989 |
| 10 | 1.3705 | 4.1504 | 3.0989 |
| 11 | 1.3 | 4.1406 | 3.0989 |
| 12 | 1.3 | 4.1406 | 3.0989 |
| 13 | 1.3 | 4.1406 | 3.0989 |
| 14 | 1.3 | 4.1406 | 3.0989 |
| 15 | 1.3 | 4.1406 | 3.0989 |
| 16 | 1.3 | 4.1406 | 3.0989 |
| 17 | 1.3 | 4.1406 | 3.0989 |
| 18 | 1.3 | 4.1406 | 3.0989 |
| 19 | 1.3 | 4.1406 | 3.0989 |
| 20 | 1.3 | 4.1406 | 3.0989 |
| 21 | 1.3 | 4.1406 | 3.0989 |
| 22 | 1.3 | 4.1406 | 3.0989 |
| 23 | 1.3 | 4.1406 | 3.0989 |
| 24 | 1.3 | 4.1406 | 3.0989 |
| 24 | 1.3 | 4.1406 | 3.0989 |

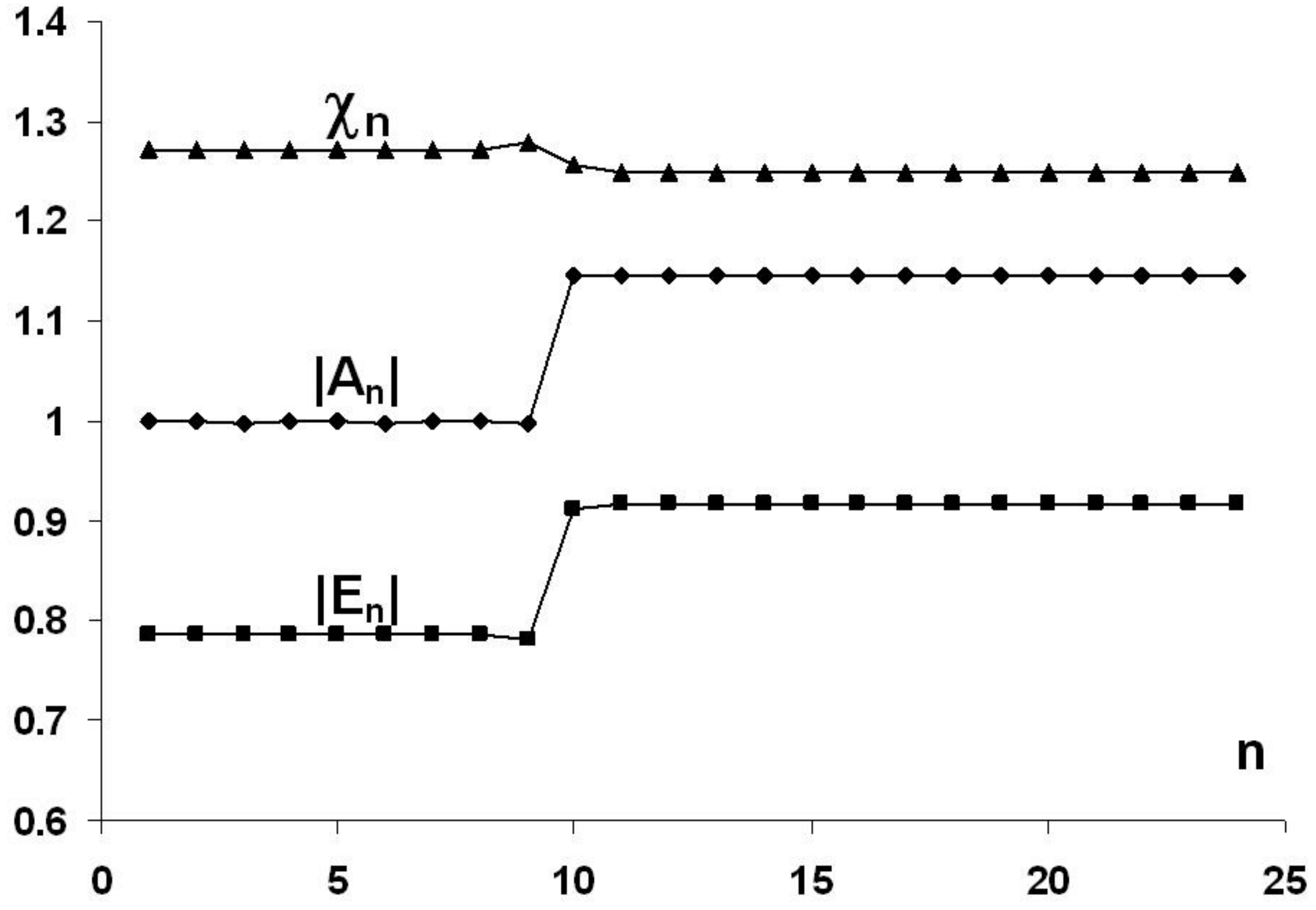

**Fig. 7 Distributions of longitudinal electric field in the cell centers ($|E_n|$), amplitudes of $E_{010}$ mode ($|A_n|$) and $\chi_n = |A_n / E_n|$**

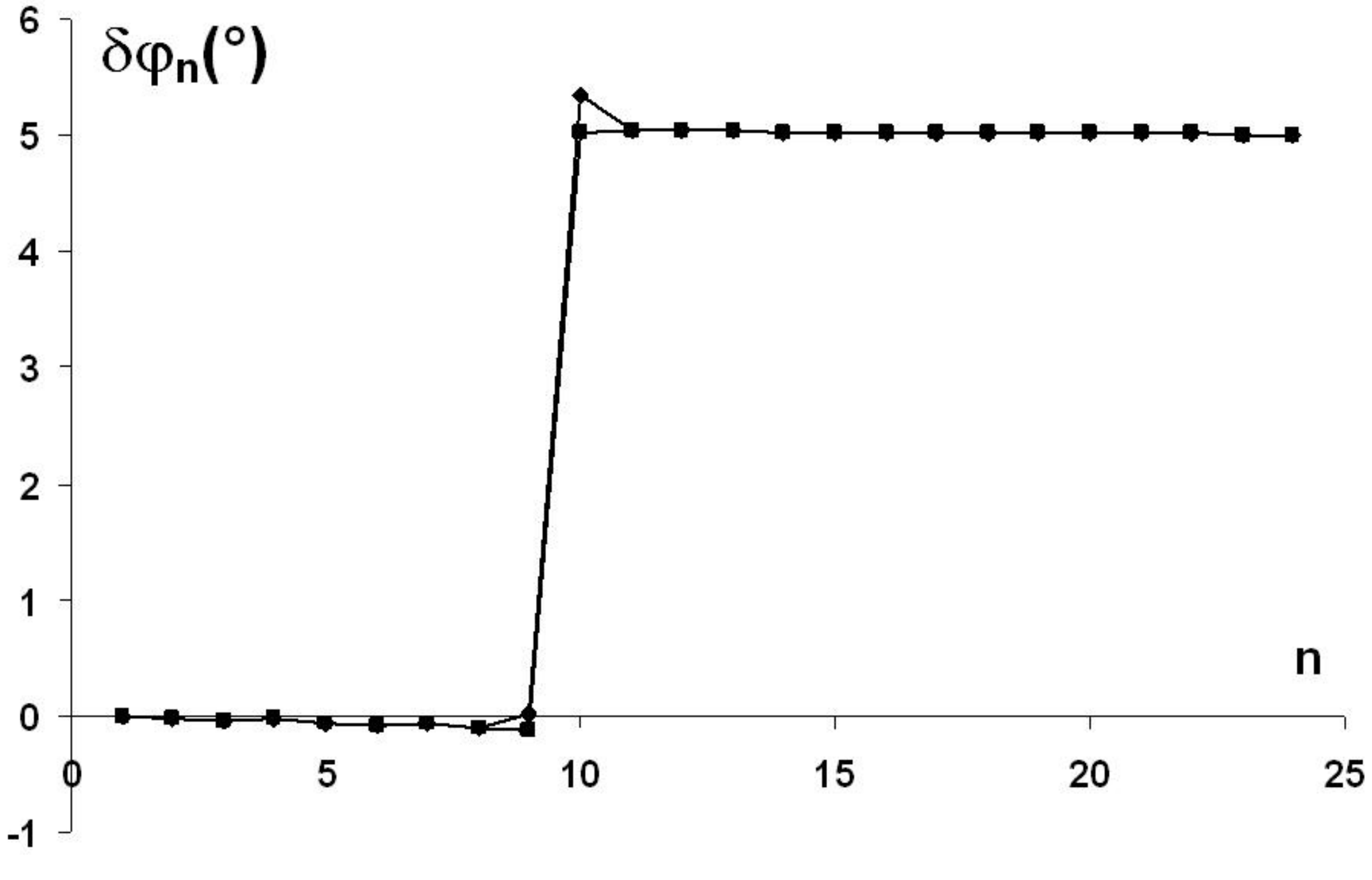

**Fig. 8 Distribution of phase deviations ($\varphi_0 = 2\pi/3$) of the longitudinal electric field in the cell centers and phase deviations of amplitude of $E_{010}$ mode**

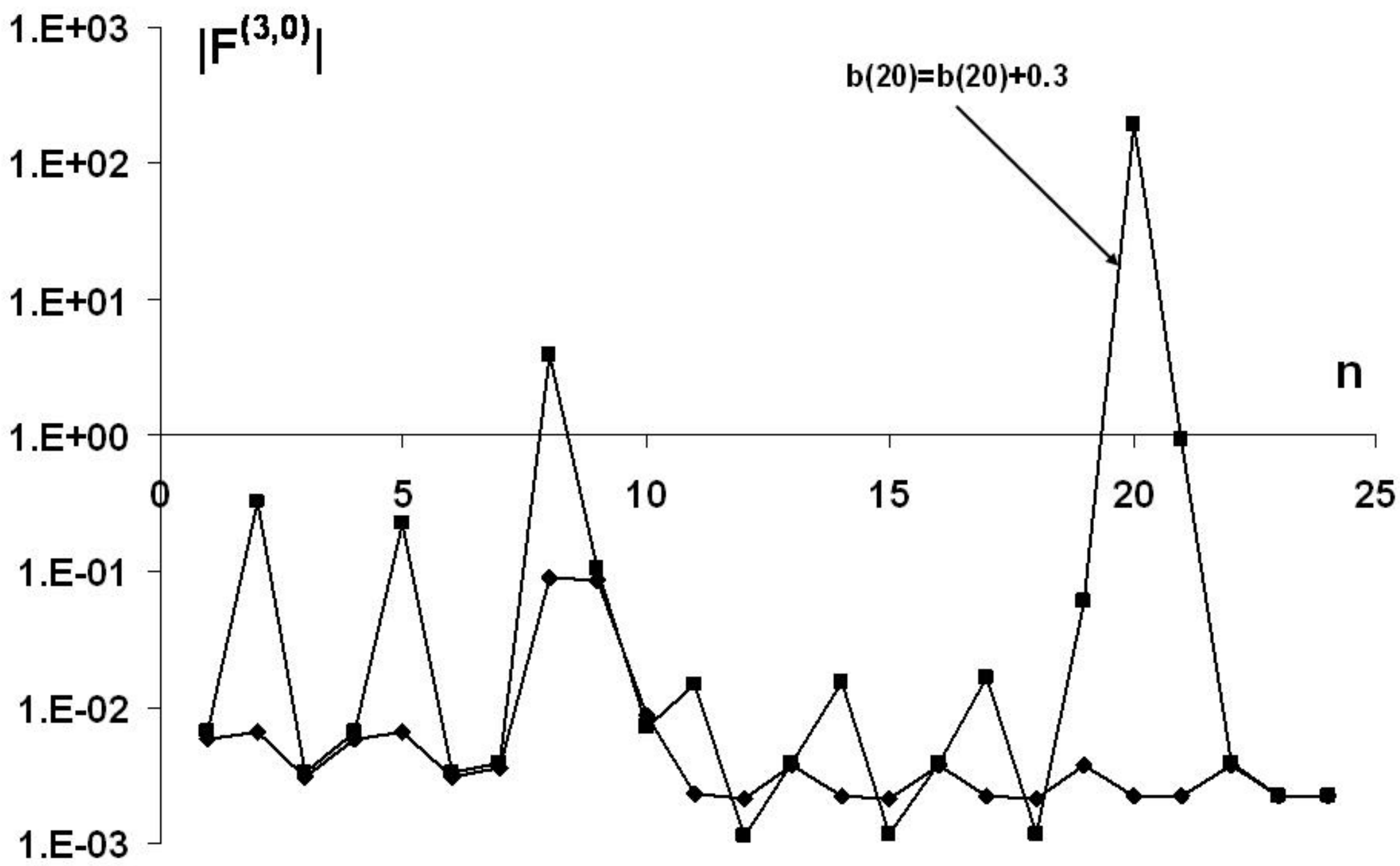

**Fig. 9 Detuning parameters that were calculated with using expression (1.39)**

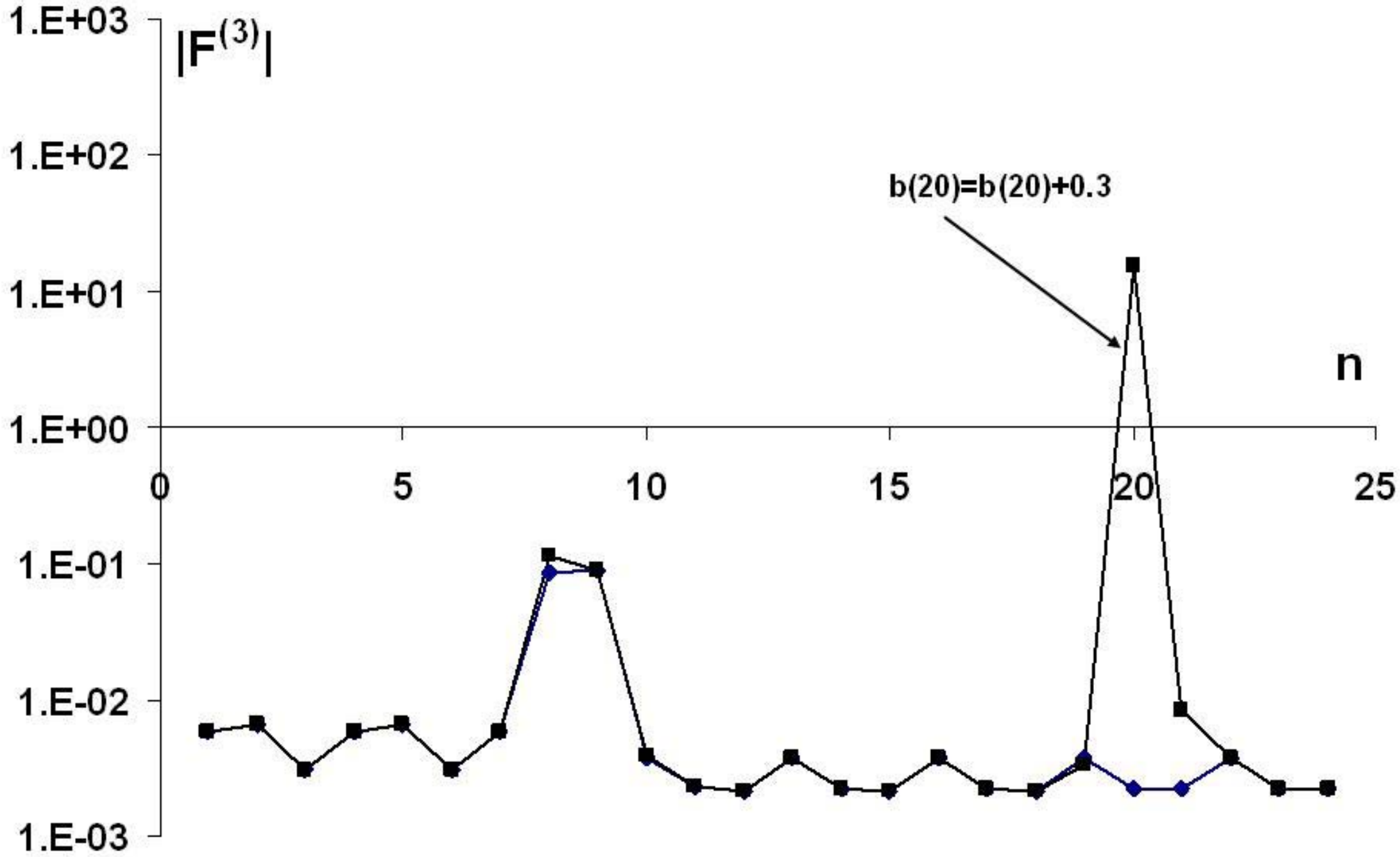

**Fig. 10 Detuning parameters that were calculated with using expression (1.35)**

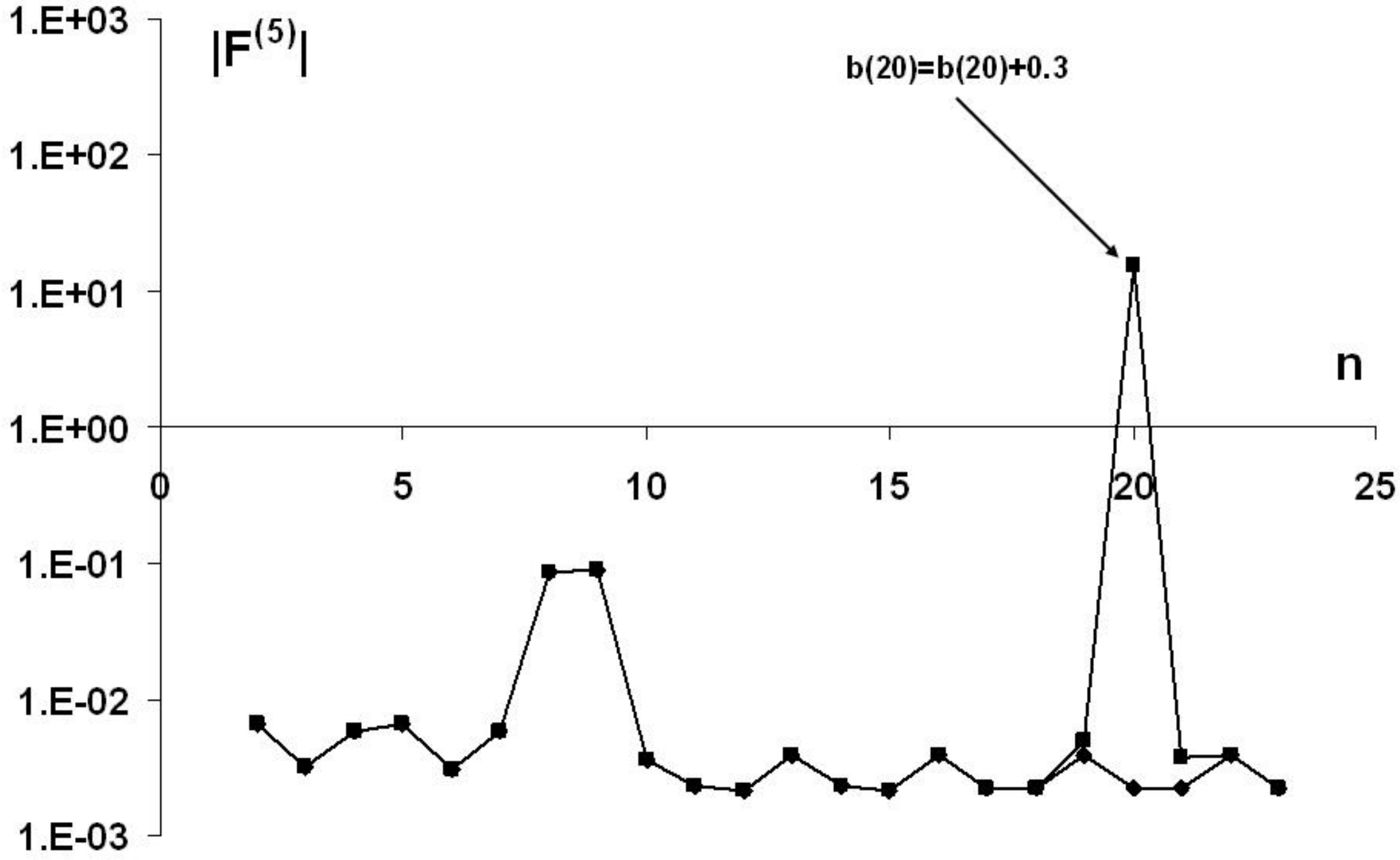

**Fig. 11 Detuning parameters that were calculated with using expression (1.44)**

**Table 3**

| N cell | a(cm) | b(cm) | d(cm) |
|---|---|---|---|
| 1 | 1.6 | 4.36528 | 1.6792 |
| 2 | 1.6 | 4.36528 | 1.6792 |
| 3 | 1.6 | 4.36528 | 1.6792 |
| 4 | 1.6 | 4.36528 | 1.6792 |
| 5 | 1.6004 | 4.3649 | 1.6792 |
| 6 | 1.6 | 4.3644 | 1.6792 |
| 7 | 1.6 | 4.3644 | 1.6792 |
| 8 | 1.6 | 4.3497 | 1.7548 |
| 9 | 1.5735 | 4.3076 | 1.8986 |
| 10 | 1.4967 | 4.2518 | 2.0978 |
| 11 | 1.4236 | 4.2154 | 2.3206 |
| 12 | 1.4 | 4.1981 | 2.5376 |
| 13 | 1.4 | 4.1875 | 2.7276 |
| 14 | 1.4 | 4.18 | 2.8715 |
| 15 | 1.4 | 4.1759 | 2.9571 |
| 16 | 1.4 | 4.1759 | 2.9571 |
| 17 | 1.4 | 4.1759 | 2.9571 |
| 18 | 1.4007 | 4.1757 | 2.9571 |
| 19 | 1.4 | 4.17524 | 2.9571 |
| 20 | 1.4 | 4.17524 | 2.9571 |
| 21 | 1.4 | 4.17524 | 2.9571 |
| 22 | 1.4 | 4.17524 | 2.9571 |
| 23 | 1.4 | 4.17524 | 2.9571 |
| 24 | 1.4 | 4.17524 | 2.9571 |
| 24 | 1.4 | 4.17524 | 2.9571 |

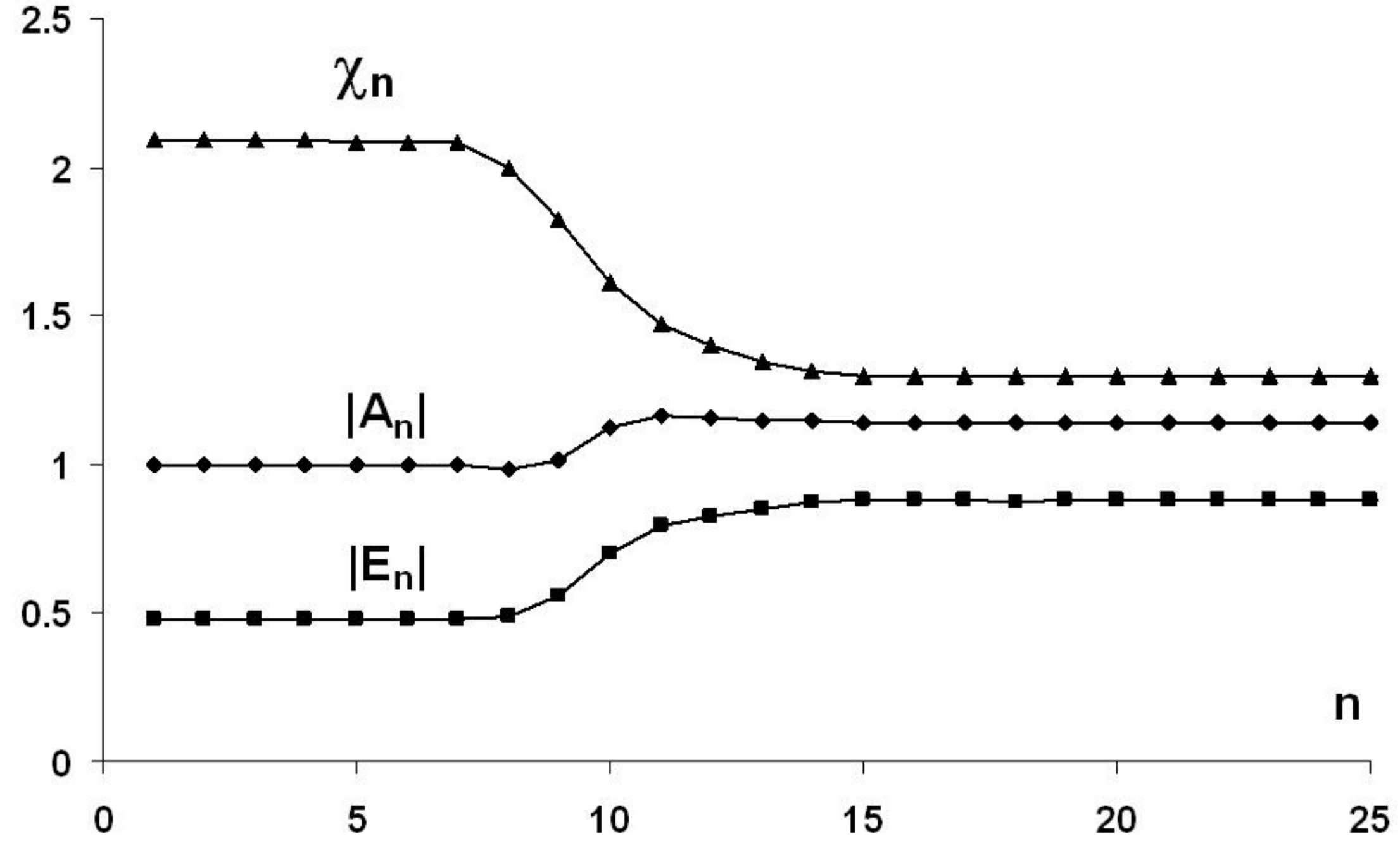

**Fig. 12 Distributions of longitudinal electric field in the cell centers ($|E_n|$), amplitudes of $E_{010}$ mode ($|A_n|$) and $\chi_n = |A_n / E_n|$**

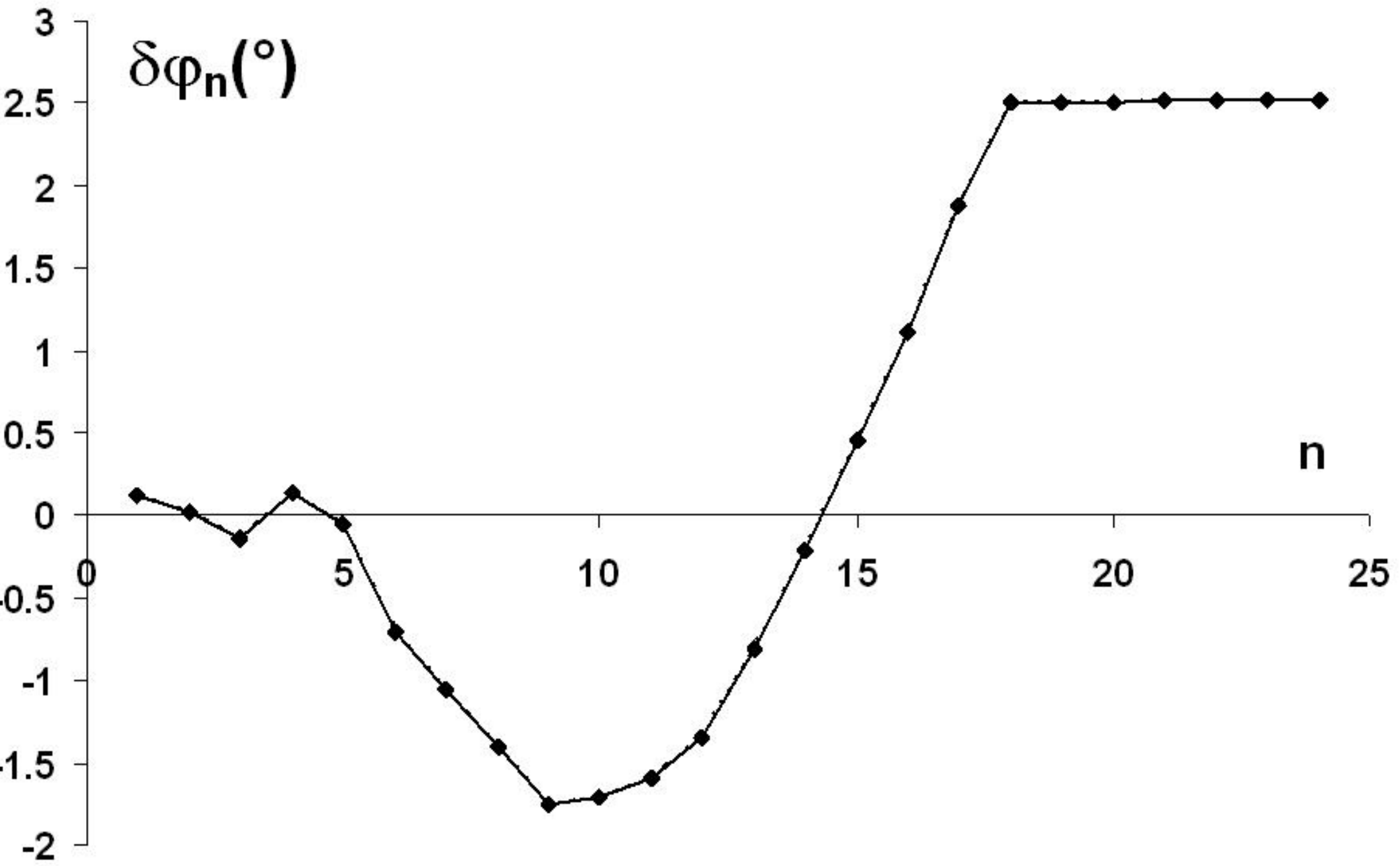

**Fig. 13 Distribution of phase deviations ($\varphi_0 = 2\pi / 3$) of the longitudinal electric field in the cell centers and phase deviations of amplitude of $E_{010}$ mode**

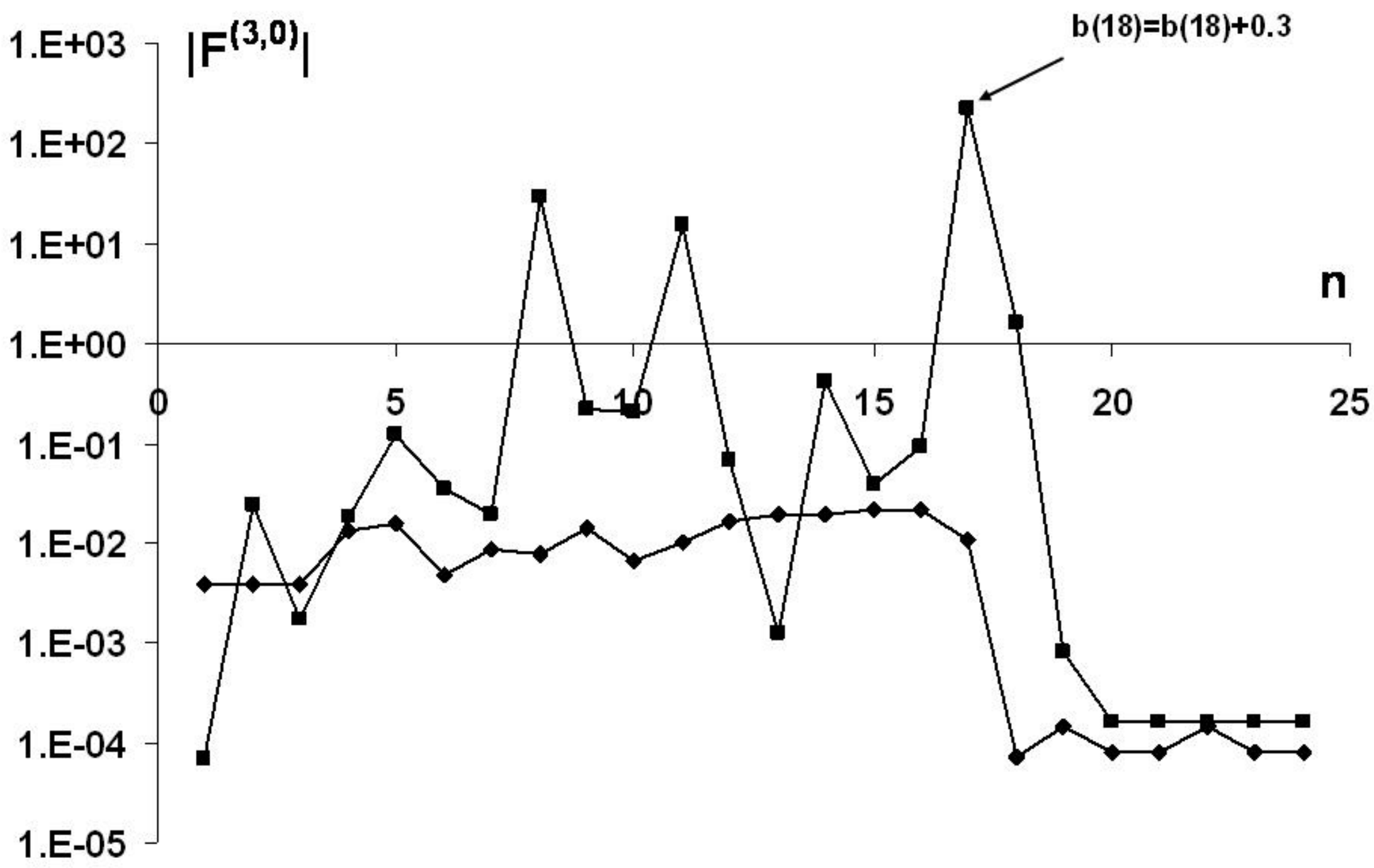

**Fig. 14 Detuning parameters that were calculated with using expression (1.39)**

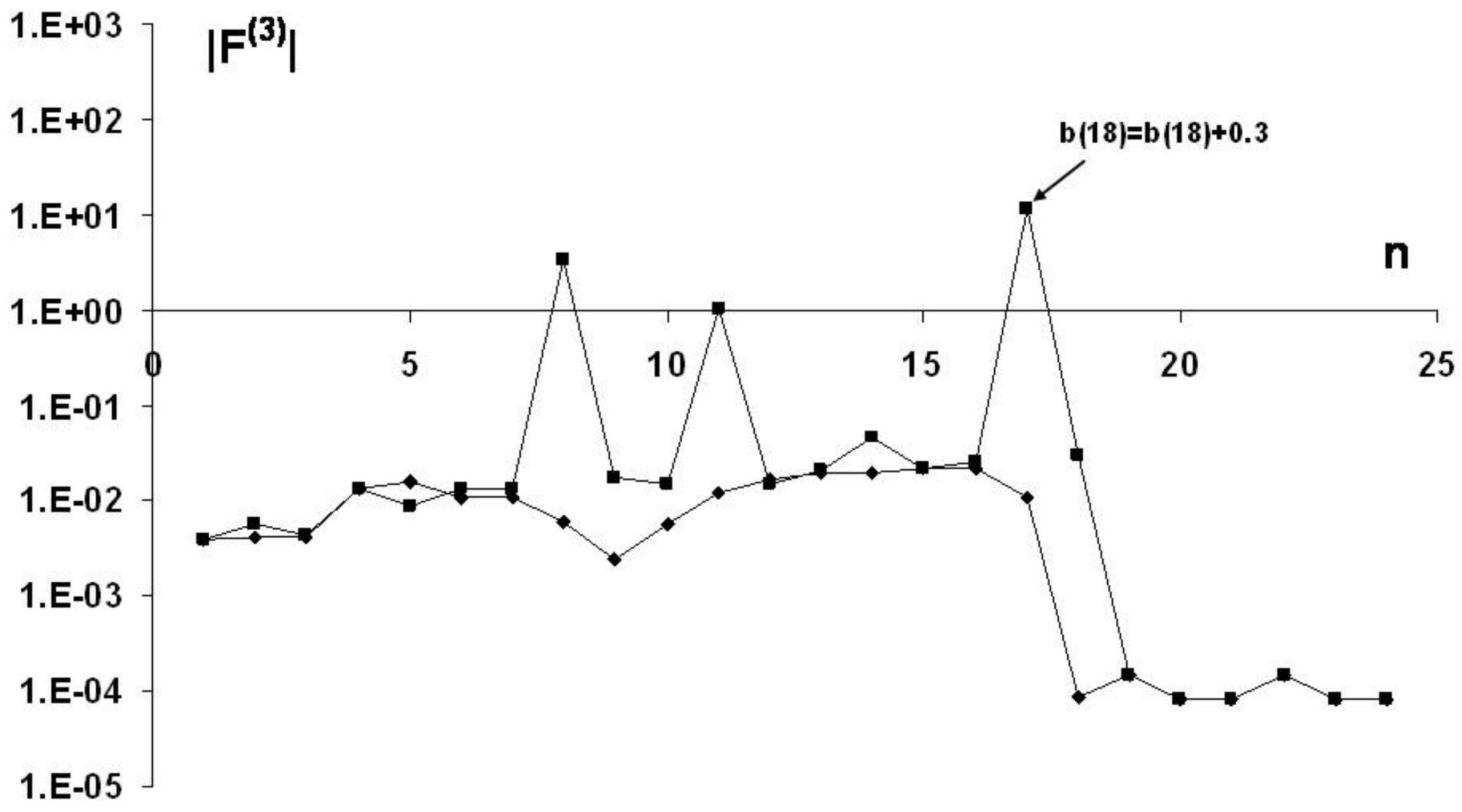

**Fig. 15 Detuning parameters that were calculated with using expression (1.35)**

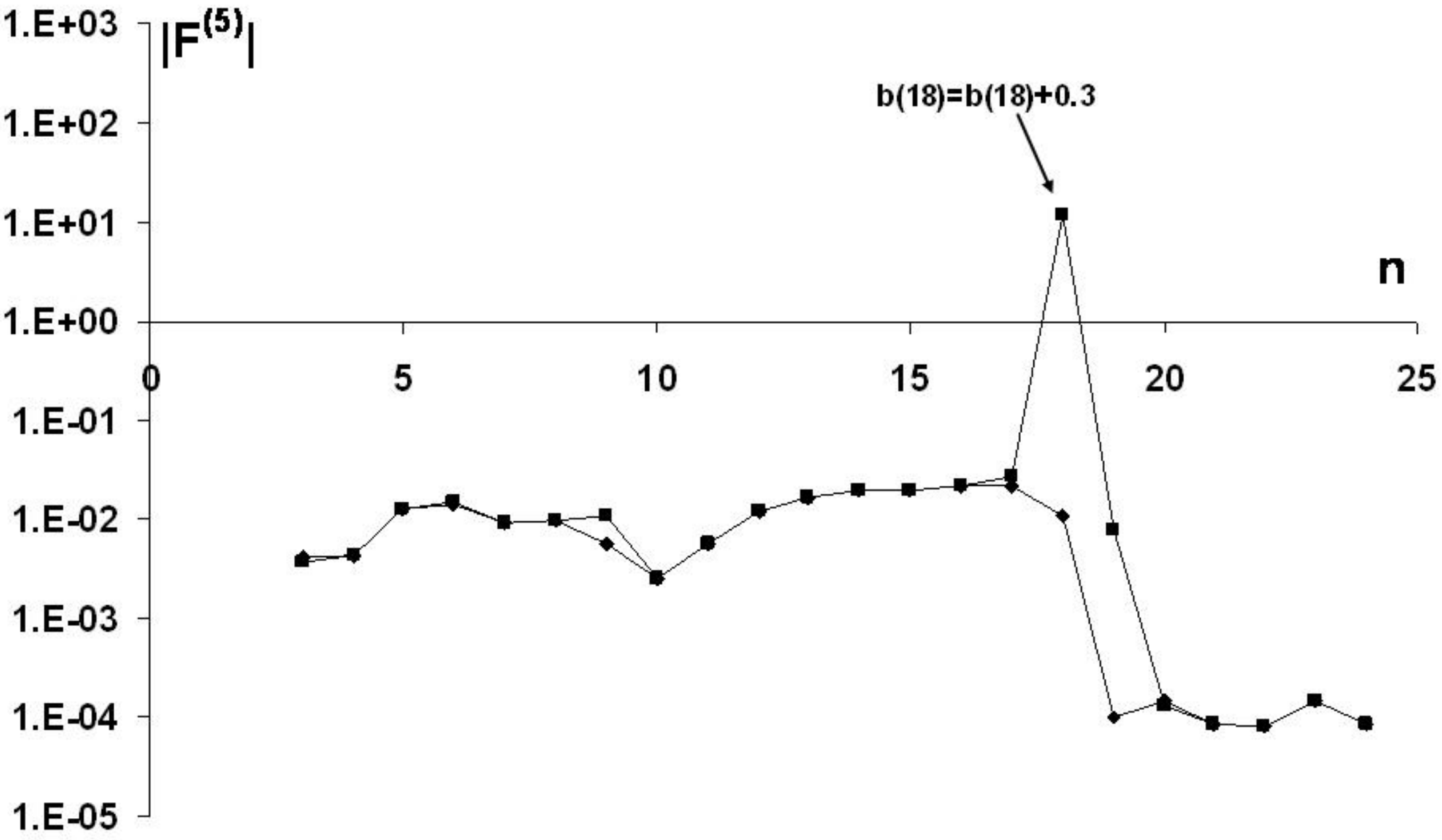

**Fig. 16 Detuning parameters that were calculated with using expression (1.44)**

## 4 Reflections in nonuniform DLW

Consider the DLW with geometry that support some distributions (1.10) with $\varphi_k^{(0)} = \varphi$ and (1.13) with $\phi_k^{(0)} \approx \varphi$. Suppose that one cell with index $s$ ($s \geq 2$) has a deviation in geometric sizes. Let us find the amplitude distribution in such DLW. The system of equations (1.7) we can rewrite in the form

$$Z_k^{(0)} \bar{A}_k = \alpha_{k,k-1}^{(0)} \bar{A}_{k-1} + \alpha_{k,k+1}^{(0)} \bar{A}_{k+1}, \ k < s. \tag{1.47}$$

$$Z_s \bar{A}_s = \alpha_{s,s-1} \bar{A}_{s-1} + \alpha_{s,s+1} \bar{A}_{s+1}. \tag{1.48}$$

$$Z_k^{(0)} \bar{A}_k = \alpha_{k,k-1}^{(0)} \bar{A}_{k-1} + \alpha_{k,k+1}^{(0)} \bar{A}_{k+1}, \ k > s. \tag{1.49}$$

The amplitude in the cell with $k = 0$ we represent in the form

$$\bar{A}_0 = A_0 + \Gamma A_0. \tag{1.50}$$

Then for other cells we can write

$$\bar{A}_k = A_0 \prod_{j=1}^{k} R_{j,j-1}^{(0)} \exp(ik\varphi) + \Gamma A_0 \prod_{j=1}^{k} \tilde{R}_{j,j-1}^{(0)} \exp(-ik\varphi), \ k < s, k > 0. \tag{1.51}$$

$$\bar{A}_k = A_0 T \prod_{j=1}^{k} R_{j,j-1}^{(0)} \exp(ik\varphi), \ k > s. \tag{1.52}$$

Substituting (1.51) and (1.52) into equations (1.47)-(1.49), we find “the reflection coefficient” $\Gamma$

$$\begin{gathered} \alpha_{s-1,s-2}^{(0)} \left\{ \frac{\exp(-i\varphi)}{R_{s-1,s-2}^{(0)}} - \frac{\exp(i\varphi)}{\tilde{R}_{s-1,s-2}^{(0)}} \right\} \Gamma = -\prod_{j=1}^{s-1} \frac{R_{j,j-1}^{(0)}}{\tilde{R}_{j,j-1}^{(0)}} \exp(2is\varphi) \times R_{s,s-1}^{(0)} \alpha_{s-1,s}^{(0)} \times \\ \times \left\{ \frac{R_{s,s-1}^{(0)}}{\alpha_{s,s-1}^{(0)}} Z_s^{(0)} F_s^{(3)} - \frac{\left(\alpha_{s,s-1} - \alpha_{s,s-1}^{(0)}\right)}{\alpha_{s,s-1}^{(0)}} \exp(-i\varphi) - \frac{\left(\alpha_{s,s+1} - \alpha_{s,s+1}^{(0)}\right)}{\alpha_{s,s-1}^{(0)}} \exp(i\varphi) R_{s+1,s}^{(0)} R_{s,s-1}^{(0)} \right\} \end{gathered}, \tag{1.53}$$

where

$$Z_s = Z_s^{(0)} + F_s^{(3)} Z_s^{(0)}. \tag{1.54}$$

Note that $F_s^{(3)}$ differs from $F_s^{(3,m)}$ (see (1.35)).

In most cases $\alpha_{k,j} \sim b_k^{-2}$ and $\omega_k \sim b_k^{-1}$

Then

$$Z_s^{(0)} F_s^{(3)} = Z_s - Z_s^{(0)} \approx \left[ -2\alpha_{s,s}^{(0)} + 2\frac{\omega^2}{\omega_s^{(0)2}} - i\frac{\omega}{\omega_s^{(0)} Q_k} \right] \frac{\Delta b_k}{b_k^{(0)}} \approx 2\frac{\Delta b_k}{b_k^{(0)}}. \quad (1.55)$$

Since $\dfrac{\left(\alpha_s - \alpha^{(0)}\right)}{\alpha^{(0)}} \approx -2\dfrac{\Delta b_s}{b_s^{(0)}}$ and

$$Z_s^{(0)} \tilde{F}_k^{(3)} = Z_k - Z_k^{(0)} = Z_s^{(m)} - Z_s^{(0)} + \left(Z_s - Z_s^{(m)}\right) =$$

$$= Z_s^{(0)} F_k^{(3,m)} + \frac{\left(\alpha_{k,k-1} - \alpha_{k,k-1}^{(0)}\right)\bar{A}_{k-1}^{(m)} + \left(\alpha_{k,k+1} - \alpha_{k,k+1}^{(0)}\right)\bar{A}_{k+1}^{(m)}}{\bar{A}_k^{(m)}} \approx Z_s^{(0)} F_k^{(3,m)}, \quad (1.56)$$

we finally obtain

$$\left\{ \frac{\exp(-i\varphi)}{R_{s-1,s-2}^{(0)}} - \frac{\exp(i\varphi)}{\tilde{R}_{s-1,s-2}^{(0)}} \right\} \Gamma_s \approx -\prod_{j=1}^{s-1} \frac{R_{j,j-1}^{(0)}}{\tilde{R}_{j,j-1}^{(0)}} \exp(2is\varphi) \frac{R_{s,s-1}^{(0)2}}{\alpha_{s-1,s-2}^{(0)}} Z_s^{(0)} F_s^{(3,m)}. \quad (1.57)$$

Note that

$$\prod_{j=1}^{s-1} R_{j,j-1}^{(0)} = \left| \frac{\bar{A}_1^{(0)}}{\bar{A}_0^{(0)}} \right| \left| \frac{\bar{A}_2^{(0)}}{\bar{A}_1^{(0)}} \right| \cdots\cdots \left| \frac{\bar{A}_{s-1}^{(0)}}{\bar{A}_{s-2}^{(0)}} \right| = \left| \frac{\bar{A}_{s-1}^{(0)}}{\bar{A}_0^{(0)}} \right| \quad (1.58)$$

For the constant-impedance structure we obtain

$$\Gamma_s \approx -i\frac{\cos\varphi}{\sin\varphi} \frac{R^{s+1}}{\tilde{R}^{s-1}} \exp(2is\varphi) F_s^{(3,m)} \approx -i\frac{\cos\varphi}{\sin\varphi} \exp(-2\mu s) \exp(2is\varphi) F_s^{(3,m)}, \quad (1.59)$$

where

$$\mu \approx \frac{1}{2\alpha \sin\varphi Q}. \quad (1.60)$$

In the case of the constant-gradient structure

$$\Gamma_s \approx -i\frac{\exp(2is\varphi)}{\sin\varphi\, \alpha_{s-1,s-2}^{(0)} \prod_{j=1}^{s-2} \tilde{R}_{j,j-1}^{(0)}} Z_s^{(0)} F_s^{(3.m)} \quad (1.61)$$

Since (see (1.29)) $\tilde{R}_{k,k-1}^{(0)} \approx \alpha_{k,k-1}^{(0)} / \alpha_{k,k+1}^{(0)}$, the product can be written in the form

$$\prod_{j=1}^{s-2} \tilde{R}_{j,j-1}^{(0)} = \frac{\alpha_{1,0}^{(0)}}{\alpha_{1,2}^{(0)}} \frac{\alpha_{2,1}^{(0)}}{\alpha_{2,3}^{(0)}} \cdots\cdot \frac{\alpha_{s-3,s-4}^{(0)}}{\alpha_{s-3,s-2}^{(0)}} \frac{\alpha_{s-2,s-3}^{(0)}}{\alpha_{s-2,s-1}^{(0)}} \approx \frac{\alpha_{1,0}^{(0)}}{\alpha_{s-2,s-1}^{(0)}}. \quad (1.62)$$

Finally

$$\Gamma_s \approx -i\frac{\exp(2is\varphi)\alpha_{s-2,s-1}^{(0)}}{\sin\varphi\, \alpha_{1,0}^{(0)} \alpha_{s-1,s-2}^{(0)}} Z_s^{(0)} F_s^{(3,m)} \approx -i\frac{\exp(2is\varphi)}{\sin\varphi\, \alpha_{1,0}^{(0)}} Z_s^{(0)} F_s^{(3,m)}. \quad (1.63)$$

Note, that expression (1.57) was obtained under assumption that $\phi_k^{(0)} \approx \varphi$ for backward field. As our consideration has shown (see (1.27)), it is needed to check this assumption for each specific case.

### Conclusions

On the base of general approach we obtain some results that can be useful in the process of tuning of nonunifrom DLWs. Our consideration has shown that simple values that characterize the detuning of the cells can be introduced only for the DLW with parameters that change very slow. In general case it is needed to conduct full numerical simulation of specific

DLW and obtain all necessary coupling coefficients. After that one can start the tuning process. The coupled cavity model can be good approach in this procedure. There is one problem in the developed coupled cavity model – taking into account the rounding of the disk hole edges [2].